# Effect of Chemical Pressure on High Temperature Ferrimagnetic Double Perovskites $Sr_2CrOsO_6$ and $Ca_2CrOsO_6$

Ryan Morrow,[1] Jennifer R. Soliz,[1,2] Adam J. Hauser,[3,4] James C. Gallagher,[3] Michael A. Susner,[5,6] Michael D. Sumption,[5] Adam A. Aczel,[7] Jiaqiang Yan,[6,8] Fengyuan Yang,[3] Patrick M. Woodward[1]

[1] Department of Chemistry and Biochemistry, The Ohio State University, Columbus, Ohio 43210-1185, USA

[2] Edgewood Chemical Biological Center, 5183 Blackhawk Road, Aberdeen Proving Ground, Maryland 21010, USA

[3] Department of Physics, The Ohio State University, Columbus, Ohio 43210-1185, USA

[4] California Nanosystems Institute, University of California, Santa Barbara, California 93106, USA

[5] Department of Materials Science and Engineering, The Ohio State University, Columbus, Ohio 43210-1185, USA

[6] Materials Science and Technology Division, Oak Ridge National Laboratory, Oak Ridge, TN 37831, USA

[7] Quantum Condensed Matter Division, Oak Ridge National Laboratory, Oak Ridge, TN 37831, USA

[8] Department of Materials Science and Engineering, The University of Tennessee, Knoxville, TN 37996, USA

## Abstract

The ordered double perovskites $Sr_2CrOsO_6$ and $Ca_2CrOsO_6$ have been synthesized and characterized with neutron powder diffraction, electrical transport measurements, and high field magnetization experiments. $Sr_2CrOsO_6$ and $Ca_2CrOsO_6$ crystallize with $R\bar{3}$ and $P2_1/n$ space group symmetry, respectively. Both materials are found to be ferrimagnetic insulators with saturation magnetizations near 0.2 $\mu_B$. $Sr_2CrOsO_6$ orders at 660 K, showing non-monotonic magnetization temperature dependence, while $Ca_2CrOsO_6$ orders at 490 K and does not show non-monotonic behavior. Evidence for a theoretically predicted canted magnetic structure in $Sr_2CrOsO_6$ is sought and not found.

## Introduction

Transition metal oxides with the perovskite structure are one of the most heavily studied classes of materials due to their ability to incorporate elements from throughout the periodic table which results in properties that span the gamut of modern functional materials. Recently, ordered double perovskites with formula $A_2BB'O_6$ and consisting of a network of corner sharing $BO_6$ and $B'O_6$ octahedra have been the focus of intense research [1] due to the occurrence of highly spin-polarized electrical transport [2-4], unusual sequences of magnetic phase transitions [5-7], and geometric frustration [8-10]. One material, a ferrimagnetic insulator $Sr_2CrOsO_6$, has the distinction of possessing the highest known Curie temperature ($T_C$) amongst double perovskites at 725 K [11]. However, despite the fact that it has been the subject of numerous theoretical studies [12-14], including one that proposes a canted magnetic structure to explain its unusual non-monotonic magnetic susceptibility temperature dependence [15], $Sr_2CrOsO_6$ has not been experimentally revisited since the initial publication [11].

Herein, we revisit the magnetic properties of $Sr_2CrOsO_6$ with neutron diffraction, electrical transport measurements, and high field magnetization experiments. Additionally, we synthesize and characterize the crystal structure and magnetic properties of $Ca_2CrOsO_6$. This compound for was prepared by Sleight and Ward in 1962, but its properties have not previously been studied [16]. The results shed light on the way that chemical pressure affects the magnetic interactions and the resulting magnetic ground states of these two compounds. Their behavior is compared and contrasted with $A_2FeOsO_6$ and $A_2CoOsO_6$ ($A$ = Sr, Ca) osmate double perovskites where chemical pressure drives a crossover from an antiferromagnetic to a ferrimagnetic ground state [17-19].

## Experimental

1.6 g powder samples were prepared by solid state reaction of stoichiometric quantities of SrO (99.9% pure, Sigma Aldrich), CaO (99.9% pure, Sigma Aldrich), $CrO_3$ (99.99% pure, Sigma Aldrich), and Os metal (99.98% pure, Alfa Aesar). Reactants were ground in an argon glove box and then placed in a high density alumina tube that was sealed in a silica ampoule (40 mL volume, 3 mm wall thickness) along with a separate vessel containing $PbO_2$ that decomposed into PbO and acted as an in-situ source of $O_2$ gas. One quarter mole excess $O_2$ gas was produced in this way in order to ensure full oxidation of the reactants. The sealed tubes were heated to 1000 °C for $Sr_2CrOsO_6$ and 950 °C for $Ca_2CrOsO_6$ and held for 48 hours in a box furnace located within a fume hood as an additional precaution. Great caution must be used as heating osmium in the presence of oxygen can produce toxic $OsO_4$ gas. Larger batch sizes or ampoules with thinner walls may produce conditions resulting in rupture of the sealed tube at elevated temperatures. The following equations were used for these reactions.

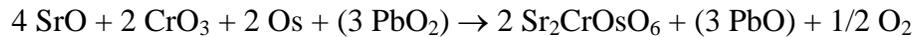
$$4\ SrO + 2\ CrO_3 + 2\ Os + (3\ PbO_2) \rightarrow 2\ Sr_2CrOsO_6 + (3\ PbO) + 1/2\ O_2$$

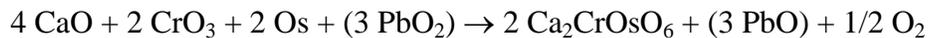
$$4\ CaO + 2\ CrO_3 + 2\ Os + (3\ PbO_2) \rightarrow 2\ Ca_2CrOsO_6 + (3\ PbO) + 1/2\ O_2$$

Laboratory X-ray diffraction measurements were conducted on a Bruker D8 Advance instrument equipped with a Ge(111) monochromator in order to verify composition and cation ordering. Time of flight neutron powder diffraction (NPD) data were collected on the POWGEN [20] beamline at Oak Ridge National Laboratory (ORNL). $Sr_2CrOsO_6$ was measured using the JANIS cryofurnace with high and low d spacing frames of 0.2760–3.0906 Å and 2.2076–10.3019 Å at 5 K and 500 K as well as single intermediate d spacing frames of 1.1038–6.1811 Å at 50 K intervals. $Ca_2CrOsO_6$ was measured in the Powgen Automatic Changer (PAC) environment at 10 K and 300 K with frames of 0.2760–4.6064 Å and 2.2076–15.3548 Å at each temperature. Rietveld refinements were conducted using the GSAS EXPGUI software package [21, 22]. Additional neutron powder diffraction data was collected on $Sr_2CrOsO_6$ using ORNL's HB-1A instrument with a constant wavelength of $\lambda = 2.367$ Å. The HB-1A sample of approximately 5 g was comprised of three combined batches of material. Diffraction patterns were collected at $T$ = 4 and 300 K using a closed cycle refrigerator, and the instrument collimation was 40'-40'-40'-80'.

$Sr_2CrOsO_6$ and $Ca_2CrOsO_6$ powders were pressed and sintered at 1000 °C and 950 °C, respectively, in evacuated sealed ampoules in order to produce bar shaped polycrystalline pellets for electrical measurements. Silver paint was applied to attach copper leads to the pellets. DC resistivity data was collected over a temperature range of 140 to 400 K for $Sr_2CrOsO_6$ and 70 to 335 K for $Ca_2CrOsO_6$ using the four-point probe method in a Quantum Design model 6000 physical property measurement system (PPMS). The samples were too resistive for accurate measurements at lower temperatures. No magnetic field was applied during the measurements. Additionally, no corrections were made for porosity.

The field dependence of the magnetization of $Sr_2CrOsO_6$ and $Ca_2CrOsO_6$ pellets was measured using the Vibrating Sample Magnetometer (VSM) option of the PPMS with the maximum field of 140 kOe at 5, 300, and 400 K for $Sr_2CrOsO_6$ and at 5 and 300 K for $Ca_2CrOsO_6$. Higher magnetic fields (± 350 kOe) were used to measure the hysteresis loops of $Sr_2CrOsO_6$ at 4.5 K and 295 K using the VSM at the National High Magnetic Field Laboratory (NHMFL).

Magnetic susceptibility of a pelletized $Sr_2CrOsO_6$ sample was obtained using a LakeShore VSM. The sample was heated to 800 K and then cooled in a 15 kOe field to field cool the sample. The field cooled susceptibility curve was then measured in field strength of 1 kOe in the temperature range from 300 to 800 K. An analogous set of measurements was conducted for $Ca_2CrOsO_6$ within the temperature range 300 to 600 K.

## Results

$Sr_2CrOsO_6$ crystallizes in the $R\bar{3}$ space group as reported previously [11], while $Ca_2CrOsO_6$ crystallizes in the monoclinic $P2_1/n$ space group common to double perovskites with a tolerance factor smaller than approximately 0.97 [23]. Cr/Os cation ordering for both compounds was determined from laboratory XRD data in order to take advantage of the substantial X-ray scattering contrast between Cr and Os. The order parameter, which is defined as $\eta = 2\theta - 1$ where $\theta$ is the occupancy of the cation on its assigned site (i.e. the Cr occupancy on the Cr-rich octahedral site), was found to be 80.2(4)% for $Sr_2CrOsO_6$ and 76.2(5)% for $Ca_2CrOsO_6$.

The results from neutron powder diffraction are given in Tables 1 and 2 for $Sr_2CrOsO_6$ and $Ca_2CrOsO_6$ respectively, while the refined NPD patterns are shown in Figures 1 and 2. The calculated bond valence sums [24] for chromium are 3.27 (5 K) and 3.38 (500 K) for $Sr_2CrOsO_6$, while they are 3.21 (10 K) and 3.22 (300 K) for $Ca_2CrOsO_6$. These results support the assignment of $Cr^{3+}$ in these materials, indicating that Os is in the 5+ oxidation state. These conclusions are supported by direct comparison to the $M-O$ bond lengths in related perovskites with these oxidation states [11, 17, 25, 26]. Other than the change in symmetry driven by the change in octahedral tilting, the most significant variance between the two is the reduction in the Cr−O−Os bond angle from 170.8° in $Sr_2CrOsO_6$ to an average of 153.2° in $Ca_2CrOsO_6$ at low temperatures. An increased bending of the $B-O-B'$ bonds is the expected response to increasing chemical pressure when the stiffness of the $B/B'-O$ bonds is higher than that of the $A-O$ bonds.

It is interesting to note that the continuous evolution of the Cr−O−Os bond angle in $Sr_2CrOsO_6$ with temperature, as shown in Figure 3, typical of rhombohedral perovskites approaching the cubic phase transition [27]. The symmetry of $Sr_2CrOsO_6$ is rhombohedral at 500 K, which is presumably quite close to a structural phase transition, given the earlier report of cubic symmetry at 540 K [11].

The electrical transport of $Sr_2CrOsO_6$ and $Ca_2CrOsO_6$ is given in Figure 4 showing linear behaviors on a $T^{-1/4}$ scale which is consistent with a variable range hopping transport model as has been reported in related double perovskite osmates [28, 29]. The resistivity of $Sr_2CrOsO_6$ increases from $8.6 \times 10^2$ $\Omega \cdot cm$ at 300 K to $2.5 \times 10^6$ $\Omega \cdot cm$ at 140 K, below which the resistance exceeds the maximum value that can be measured with our instrument, clearly demonstrating insulating behavior. A room temperature resistivity value of 10 $\Omega \cdot cm$ had been reported previously, but the temperature dependence has not been reported previously [11]. $Ca_2CrOsO_6$ also demonstrates insulating behavior, albeit with resistivity which is significantly reduced from $Sr_2CrOsO_6$.

The temperature dependence of the magnetic susceptibilities of $Sr_2CrOsO_6$ and $Ca_2CrOsO_6$ are plotted in Figure 5. The observed $T_C$ of 660 K for $Sr_2CrOsO_6$ is smaller than the value of 725 K reported previously [11] but still larger than any other known double perovskite. The difference in ordering temperature is unclear but can likely be attributed to differences in cation order, as Krockenberger *et al*. indicate full Cr/Os cation order [11]. Reductions in cation order have been shown to result in reduction in magnetic ordering temperature in other double perovskites [30]. The magnetic susceptibility is shown to increase non-monotonically from 300 K to approximately 500 K where it peaks. The temperature dependence of the magnetic susceptibility of $Ca_2CrOsO_6$ was measured between 300 and 600 K as shown in Figure 5b revealing a $T_C$ of approximately 490 K. Unlike $Sr_2CrOsO_6$, $Ca_2CrOsO_6$ does not exhibit a non-monotonic temperature dependence of the field cooled susceptibility.

In rare earth (*RE*) chromate perovskites, $RECrO_3$, a reduction in Cr−O−Cr bond angles (i.e. increased octahedral tilting) is driven by progressively smaller *RE* cations. This weakens the superexchange interactions leading to a reduction in the Neél temperature ($T_N$): $LaCrO_3$ $T_N = 282$ K, $NdCrO_3$ $T_N = 224$ K, $TbCrO_3$ $T_N = 158$ K, $YCrO_3$ $T_N = 141$ K, $LuCrO_3$ $T_N = 112$ K [31]. We observe analogous behavior in $Sr_2CrOsO_6$ and $Ca_2CrOsO_6$. This implies that the nearest neighbor antiferromagnetic coupling is the dominant exchange interaction in $A_2CrOsO_6$ double perovskites just as it is in $RECrO_3$ perovskites. The higher ordering temperatures seen in the $A_2CrOsO_6$ double perovskites imply stronger superexchange coupling than in $RECrO_3$ perovskites, clear evidence that superexchange coupling involving the $t_{2g}$ orbitals of 3d and 5d ions can be very strong.

The field dependence of the magnetization of $Sr_2CrOsO_6$ is shown in Figure 6. A saturation magnetization ($M_{sat}$) of 0.22 $\mu_B$ per formula unit (f.u.) is measured at 4.5 and 295 K under maximum applied fields of 350 kOe, while lesser fields of the available PPMS were insufficient to saturate the magnetization [11], as shown in Figure 6b. The field dependence of the magnetization of $Ca_2CrOsO_6$ at 5 K and 300 K under maximum applied fields of 140 kOe is given in Figure 6c. The magnetization clearly does not saturate under these conditions, and further high field measurements would be necessary to reach saturation. However, as the magnetization of $Ca_2CrOsO_6$ approaches similar values as $Sr_2CrOsO_6$, it can safely be assumed that both compounds are ferrimagnetic with similar saturation magnetization values.

The magnetic scattering intensity observed in the neutron powder diffraction patterns could be modeled using a collinear ferrimagnetic model. Unfortunately strong correlations between the moments on the Cr and Os sites make it difficult to independently refine the moments of the two magnetic ions. It is possible to produce similar quality fits to the data with a continuum of Cr/Os moments, provided the difference between the Cr and Os moments stays roughly constant, as has been reported in prior studies of ferrimagnetic double perovskites [17, 32].

The data can be modeled by refining the moment on chromium and neglecting any contribution from osmium. In the case of $Sr_2CrOsO_6$ this approach results in chromium moments of 2.24(6) and 4.01(8) $\mu_B$ at 500 and 5 K respectively, the latter value is physically unreasonable for a $d^3$ ion like $Cr^{3+}$. In order to systematically refine the magnetic structure with moments on both ions the moment on Cr was fixed to 2.5 $\mu_B$, in line with values recorded in earlier studies of $RE$CrO$_3$ perovskites [31], while the Os moment was allowed to refine. This resulted in an Os moment of 0.82(9) $\mu_B$ at 5 K. This value is reduced from the values obtained for $Os^{5+}$ in other studies of osmate perovskites [33-35], possibly due to Cr/Os antisite disorder. Figure 7 shows the temperature dependence of the refined Os moment using this approach. The refined moment is fairly constant in the temperature range 5 to 350 K, above which it drops substantially at 400 and 450 K. Finally at 500 K the Os moment seems to go to zero and the Cr moment reduces to 2.24(6) $\mu_B$. This is near the temperature where a maximum is seen in the magnetization measurement. It should be noted that while this refinement approach could also be consistent with both Cr and Os moments decreasing with temperature at a similar rate, that scenario would not lead to the increase in magnetization seen in Figure 5a for $Sr_2CrOsO_6$ between 300 and 500 K.

A similar approach was employed in the refinement of the magnetic moments in $Ca_2CrOsO_6$. Refining only Cr moments resulted in physically unreasonable values of 4.07(6) and 4.34(5) $\mu_B$ at 300 and 10 K, respectively. Fixing the Cr moment at 2.5 $\mu_B$ antiparallel to Os resulted in refined Os moments of 1.26(6) and 0.55(6) $\mu_B$ at 10 and 300 K respectively.

Additional purely magnetic reflections were not observed in neutron powder diffraction for either $Ca_2CrOsO_6$ or $Sr_2CrOsO_6$ precluding the consideration of canted magnetic models. As shown in Figure 8, $Sr_2CrOsO_6$ was further scrutinized with neutron scattering at Oak Ridge National Laboratory's HFIR facility on the HB-1A instrument to search for possible weak magnetic reflections in accordance with the theoretical canted magnetic structure model [15]. HB-1A is the ideal instrument for this particular investigation because of its excellent signal-to-noise ratio, arising from the combined use of a double-bounce monochromator and an analyzer. However, all weak features in the data were accounted for in powder X-ray diffraction data indicating that they are not of magnetic origin. Specifically, the strong magnetic reflection predicted to occur at $Q = 1.14$ Å$^{-1}$, corresponding to a (110) reflection on a cubic double perovskite lattice (used in the theoretical model [15]), is not observed as highlighted in Figure 8c. Thus, within the limitations of a powder neutron experiment, there is no evidence of the theoretically predicted spin canting in the magnetic structure of $Sr_2CrOsO_6$.

Perovskites containing the d$^3$ electronic configuration are known for having high magnetic transition temperatures. Examples include the antiferromagnets $NaOsO_3$ ($T_N$ = 410 K [35]), $SrRu_2O_6$ ($T_N$ = 565 K [36]), and $SrTcO_3$ ($T_N$ = 1000 K [37]). While $Sr_2CrOsO_6$ has a similar electronic configuration (3d$^3$–5d$^3$), it is ferrimagnetic rather than antiferromagnetic due to a small but non-negligible orbital contribution [33] that reduces the moment on Os preventing fully compensated magnetism between Cr and Os sublattices. Recent work suggests that spin-orbit coupling (SOC) is allowed in the 5d$^3$ case due to an electronic state which is intermediate between L-S and j-j coupling schemes, rendering the assumption that SOC is completely quenched invalid [38]. While no evidence was found for a canted magnetic structure, an alternative cause of net magnetization as proposed by Meetei *et al*. [15], for the non-monotonic magnetic susceptibility of $Sr_2CrOsO_6$ is confirmed. The decrease in magnetic scattering, observed upon warming above 350 K and modeled here as a decrease in the Os moment, is coincident with the non-monotonic portion of the magnetization temperature dependence and lends support to the idea that the magnetization of the Os and Cr sublattices have differing temperature dependences. While the neutron data have been modeled in accordance with this idea, conclusive evidence will require an elemental probe such as X-ray magnetic circular dichroism (XMCD). Unfortunately such an experiment would be very challenging due to the wide temperature range of the measurement and the exceptionally high magnetic fields needed to saturate the magnetization of the material.

## Conclusion

$Sr_2CrOsO_6$ and $Ca_2CrOsO_6$ have been synthesized and characterized by a number of techniques. The compounds form partially ordered double perovskites with $R\bar{3}$ ($Sr_2CrOsO_6$) and $P2_1/n$ ($Ca_2CrOsO_6$) space group symmetries. $Sr_2CrOsO_6$ is an insulating ferrimagnet with $T_C$ = 660 K and $M_{sat}$ = 0.22 μ$_B$/f.u. while $Ca_2CrOsO_6$ is also an insulating ferrimagnet with a reduced $T_C$ of 490 K and an $M_{sat}$ of approximately 0.2 μ$_B$/f.u. The temperature dependence of the magnetic susceptibility of $Sr_2CrOsO_6$ is non-monotonic, an unusual feature for a ferrimagnetic double perovskite. This behavior is attributed to different temperature dependences in the magnetization of the chromium and osmium sublattices. Interestingly, similar behavior is not observed for $Ca_2CrOsO_6$.

## Author Information

**Corresponding Author**

woodward@chemistry.ohio-state.edu

**Notes**

The authors declare no competing financial interest.

## Acknowledgements

Support for this research was provided by the Center for Emergent Materials an NSF Materials Research Science and Engineering Center (DMR-1420451). Additional support was provided by the U.S. Department of Energy, Office of High Energy Physics under Grant Number DE-FG02-95ER40900 and DE-SC0001304 (magnetic characterization). A portion of this work was performed at the National High Magnetic Field Laboratory, which is supported by National Science Foundation Cooperative Agreement No. DMR-1157490 and the State of Florida. A portion of this research was carried out at Oak Ridge National Laboratory's Spallation Neutron Source and High Flux Isotope Reactor, which is sponsored by the U.S. Department of Energy, Office of Basic Energy Sciences. The authors thankfully acknowledge Ashfia Huq and Pamela Whitfield for experimental assistance with POWGEN data collection.

## References

1) S. Vasala, and M. Karppinen, Prog. Solid State Chem. **43**, 1-36 (2014).

2) K.-I Kobayashi, T. Kimura, H. Sawada, K. Terakura, and Y. Tokura, Nature. **395**, 677-680 (1998).

3) K.-I. Kobayashi, T. Kimura, Y. Tomioka, H. Sawada, K. Terakura, and Y. Tokura, Phys. Rev. B **59**, 11159-11162 (1999).

4) A. J. Hauser, J. R. Soliz, M. Dixit, R. E. A. Williams, M. A. Susner, B. Peters, L. M. Mier, T. L. Gustafson, M. D. Sumption, H. L. Fraser, P. M. Woodward, and F. Y. Yang, Phys. Rev. B **85**, 161201(R) (2012).

5) R. Morrow, R. Mishra, O. R. Restrepo, M. R. Ball, W. Windl, S. Wurmehl, U. Stockert, B. Büchner, and P. M. Woodward, J. Am. Chem. Soc. **135**, 18824-18830 (2013).

6) A. K. Paul, M. Reehuis, V. Ksenofontov, B. Yan, A. Hoser, D. M. Tobbens, P. Adler, M. Jansen, and C. Felser, Phys. Rev. Lett. **111**, 167205 (2013).

7) B. Yan, A. K. Paul, S. Kanungo, M. Reehuis, A. Hoser, D. M. Többens, W. Schnelle, R. C. Williams, T. Lancaster, F. Xiao, J. S. Möller, S. J. Blundell, W. Hayes, C. Felser, and M. Jansen, Phys. Rev. Lett. **112**, 147202 (2014).

8) A. A. Aczel, P. J. Baker, D. E. Bugaris, J. Yeon, H.-C. zur Loye, T. Guidi, and D. T. Adroja, Phys. Rev. Lett. **112**, 117603 (2014).

9) T. Aharen, J. E. Greedan C. A. Bridges, A. A. Aczel, J. Rodriguez, G. MacDougall, G. M. Luke, T. Imai, V. K. Michaelis, S. Kroeker, H. Zhou, C. R. Wiebe, and L. M. D. Cranswick, Phys. Rev. B **81**, 224409 (2010).

10) C. R. Wiebe, J. E. Greedan, P. P. Kyriakou, G. M. Luke, J. S, Gardner, I. M. Gat-Malureanu, P. L. Russo, A. T. Savici, and Y. J. Uemura, Phys. Rev. B, **68**,134410 (2003).

11) Y. Krockenberger, K. Mogare, M. Reehuis, M. Tovar, M. Jansen, G. Vaitheeswaran, V. Kanchana, F. Bultmark, A. Delin, F. Wilhelm, A. Rogalev, A. Winkler, and L. Alff, Phys. Rev. B. **75**, 020404(R) (2007).

12) T. K. Mandal, C. Felser, M. Greenblatt, and J. Kübler. Phys. Rev. B **78**, 134431 (2008).

13) H. Das, P. Snayal, T. Saha-Dasgupta, and D. D. Sarma, Phys. Rev. B **83**, 104418 (2011).

14) K.-W. Lee and W. E. Pickett, Phys. Rev. B **77**, 115101 (2008).

15) O. N. Meetei, O. Erten, M. Randeria, N. Trivedi, and P. M. Woodward, Phys. Rev. Lett. **110**, 087203 (2013).

16) A. W. Sleight, J. Longo, and R. Ward, Inorg. Chem. **1**, 245–250 (1962).

17) R. Morrow, J. W. Freeland, and P. M. Woodward, Inorg. Chem. **53**, 7983-7992 (2014).

18) H. L. Feng, M. Arai, Y. Matsushita, Y. Tsujimoto, Y. Guo, C. I. Sathish, X. Wang, Y. Yuan, M. Tanaka, and K. J. Yamaura. J. Am. Chem. Soc. **136**, 3326–3329 (2014).

19) R. Morrow, J.-Q. Yan, M. A. McGuire, J. W. Freeland, D. Haskel, and P. M. Woodward, Phys Rev B (2015, accepted) arXiv:1503.00029v2.

20) A. Huq, J. P. Hodges O. Gourdon, and L. Heroux. Zeitschrift für Kristallographie Proceedings, **1**, 127-135 (2011).

21) A. C. Larson and R. B. Von Dreele, Los Alamos National Laboratory Report LAUR 86-748 2000.

22) B. H. Toby, EXPGUI, a graphical user interface for GSAS. J. Appl. Cryst. **34**, 210-213 (1991).

23) M. W. Lufaso, P. W. Barnes, and P. M. Woodward, Acta Cryst. B **62**, 397-410 (2006).

24) N. E. Brese and M. O'Keefe, Acta Cryst. B **47**, 192-197 (1991).

25) N. Sakaia, H. Fjellvâga, and B. C. Haubackb, J. Solid State Chem. **121**, 202–213 (1996).

26) Y. G. Shi, Y. F. Guo, S. Yu, M. Arai, A. A. Belik, A. Sato, K. Yamaura, E. Takayama-Muromachi, H. F. Tian, H. X. Yang, J. Q. Li, T. Varga, J. F. Mitchell, and S. Okamoto, Phys. Rev. B **80**, 161104(R) (2009).

27) C. J. Howard, K. J. Kennedy, and B. C. Chakoumakos, J Phys.: Condens. Matter **12**, 349-365 (2000).

28) A. K. Paul, M. Reehuis, C. Felser, P. M. Abdala, and M. Jansen, Z. Anorg. Allg. Chem. **639**, 2421–2425 (2013).

29) H. L. Feng, Y. Tsujimoto, Y. Guo, Y. Sun, C. I. Sathish, and K. Yamaura, High Pressure Res. **33**, 221-228 (2013).

30) A. S. Ogale, S. B. Ogale, R. Ramesh, and T. Venkatesan, Appl. Phys. Lett. **75**, 537 (1999).

31) D. E. Cox, IEEE Trans. Magn. **8**, 161 (1972).

32) C. M. Thompson, L. Chi, A. M. Hayes, M. N. Wilson, T. J. S. Munsie, I. P. Swainson, A. P. Grosvenor, G. M. Luke, J. E. Greedan, Dalton Transactions **44**, 10806-10816 (2015).

33) A. E. Taylor, R. Morrow, D. J. Singh, S. Calder, M. D. Lumsden, P. M. Woodward and A. D. Christianson, Phys. Rev. B **91**, 100406(R) (2015).

34) A. K. Paul, A. Sarapulova, P. Adler, M. Reehuis, S. Kanungo, D. Mikhailova, W. Schnelle, Z. Hu, C. Kuo, V. Siruguri, S. Rayaprol, Y. Soo, B. Yan, C. Felser, L. H. Tjeng, and M. Jansen, Z. Anorg. Allg. Chem. **641**, 197 (2015).

35) S. Calder, V. O. Garlea, D. F. McMorrow, M. D. Lumsden, M. B. Stone, J. C. Lang, J.-W. Kim, J. A. Schlueter, Y. G. Shi, K. Yamaura, Y. S. Sun, Y. Tsujimoto, and A. D. Christianson, Phys. Rev. Lett. **108**, 257209 (2012).

36) W. Tian, C. Svoboda, M. Ochi, M. Matsuda, H. B. Cao, J.-G. Cheng, B. C. Sales, D. G. Mandrus, R. A. Arita, N. Trivedi, and J.-Q. Yan. Phys. Rev. Lett. (in review) arXiv:1504.03642.

37) E. E. Rodriguez, F. Poineau, A. Llobet, B. J. Kennedy, M. Avdeev, G. J. Thorogood, M. L. Carter, R. Seshadri, D. J. Singh, and A. K. Cheetham, Phys. Rev. Lett. **106**, 067201 (2011).

38) H. Matsuura and K. Miyake, J. Phys. Soc. Jpn. **82**, 073703 (2013).

| Temperature (K) | 5 | 500 |
|---|---|---|
| Space Group | $R\bar{3}$ | $R\bar{3}$ |
| $a$ (Å) | 5.5200(1) | 5.5300(1) |
| $c$ (Å) | 13.4403(3) | 13.5242(6) |
| $V$ (Å)$^3$ | 354.662) | 358.2(1) |
| $R_{wp}$ | 2.74% | 3.60% |
| | | |
| Cr−O (×6, Å) | 1.964(2) | 1.952(1) |
| Os−O (×6, Å) | 1.944(2) | 1.958(1) |
| ∠Cr−O−Os (°) | 170.82(5) | 176.5(1) |
| | | |
| Sr $z$ | 0.2500(3) | 0.2502(7) |
| O $x$ | 0.3350(6) | 0.3323(8) |
| O $y$ | 0.1958(3) | 0.1770(6) |
| O $z$ | 0.4162(2) | 0.4166(4) |

TABLE 1. Neutron powder diffraction parameters obtained from Rietveld refinement for $Sr_2CrOsO_6$ at 5 and 500K.

| Temperature (K) | 10 | 300 |
|---|---|---|
| Space Group | $P2_1/n$ | $P2_1/n$ |
| a (Å) | 5.3513(1) | 5.36287(9) |
| b (Å) | 5.4561(1) | 5.4541 (1) |
| c (Å) | 7.6204(1) | 7.6321(1) |
| V (Å)3 | 222.50(1) | 223.24 (1) |
| β (°) | 90.092(2) | 90.074(2) |
| $R_{wp}$ | 3.98% | 3.89% |
| | | |
| Cr−O1 (×2, Å) | 1.973(3) | 1.971(2) |
| Cr−O2 (×2, Å) | 1.973(2) | 1.970(2) |
| Cr−O3 (×2, Å) | 1.967(3) | 1.972(3) |
| | | |
| Os−O1 (×2, Å) | 1.955(3) | 1.959(2) |
| Os−O2 (×2, Å) | 1.960(2) | 1.962(2) |
| Os−O3 (×2, Å) | 1.946(3) | 1.943(3) |
| | | |
| ∠Cr−O1−Os (°) | 153.3(2) | 153.4(2) |
| ∠Cr−O2−Os (°) | 152.6(2) | 153.1(2) |
| ∠Cr−O3−Os (°) | 153.8(1) | 154.20(8) |
| | | |
| Ca x | −0.0096(3) | −0.0086(4) |
| Ca y | 0.0458(2) | 0.0434(2) |
| Ca z | 0.2499(4) | 0.2505(5) |
| O1 x | 0.2044(4) | 0.2050(4) |
| O1 y | 0.2087(4) | 0.2081(5) |
| O1 z | −0.0407(3) | −0.0403(4) |
| O2 x | 0.2067(3) | 0.2069(4) |
| O2 y | 0.2064(4) | 0.2073(5) |
| O2 z | 0.5428(3) | 0.5417(4) |
| O3 x | 0.4196(2) | 0.4211(3) |
| O3 y | −0.0200(2) | −0.0199(3) |
| O3 z | 0.2514(3) | 0.2519(4) |

TABLE 2: Neutron powder diffraction parameters obtained from Rietveld refinement for $Ca_2CrOsO_6$ at 10 and 300 K.

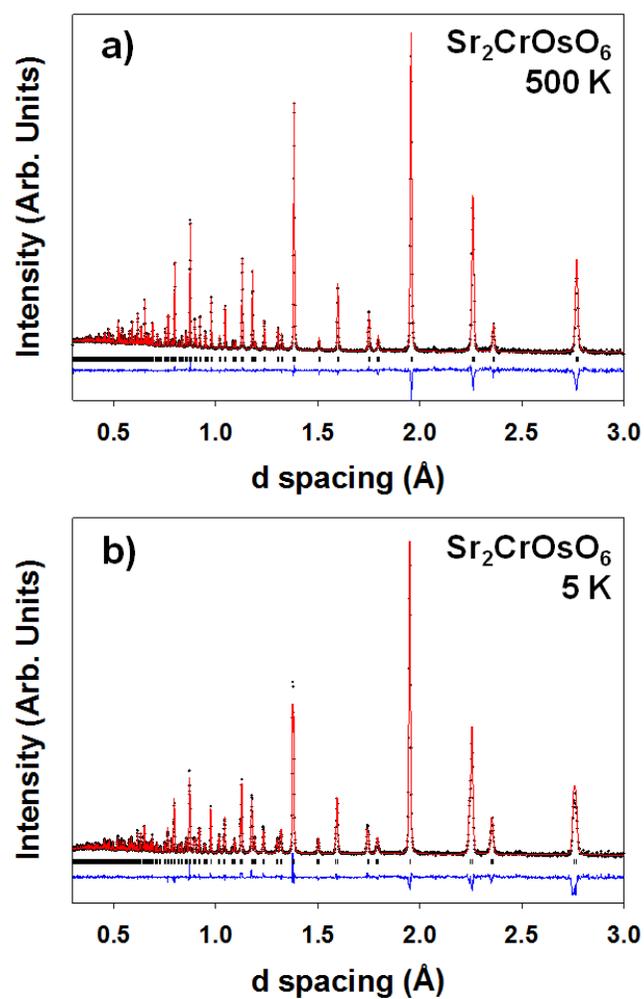

FIG. 1 (color online). Refined neutron powder diffraction pattern of $Sr_2CrOsO_6$ at a) 500 K and b) 5 K. Black symbols, red curves, and blue curves correspond to observed data, the calculated patterns, and the difference curve, respectively. The black hashes correspond to both the nuclear and magnetic peak positions of the compound.

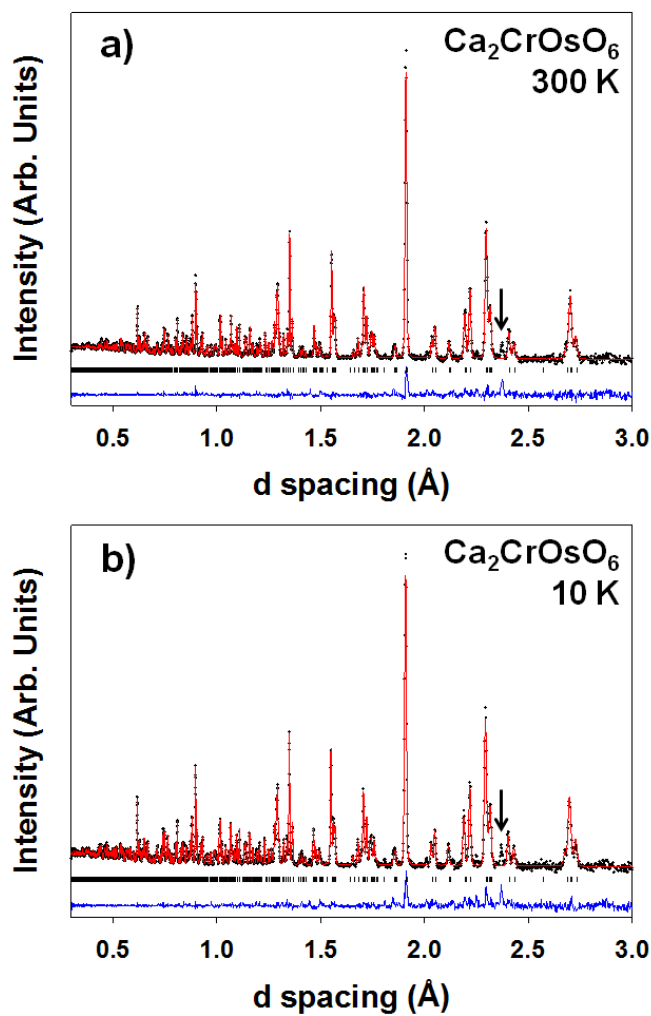

FIG. 2 (color online). Refined neutron powder diffraction pattern of $Ca_2CrOsO_6$ at a) 300 K and b) 10 K. Black symbols, red curves, and blue curves correspond to observed data, the calculated patterns, and the difference curves, respectively. Black hashes correspond to both the nuclear and magnetic peak positions of the compound. There is a trace amount of CaO which is unrefined in the patterns with a single peak identified by the black arrow.

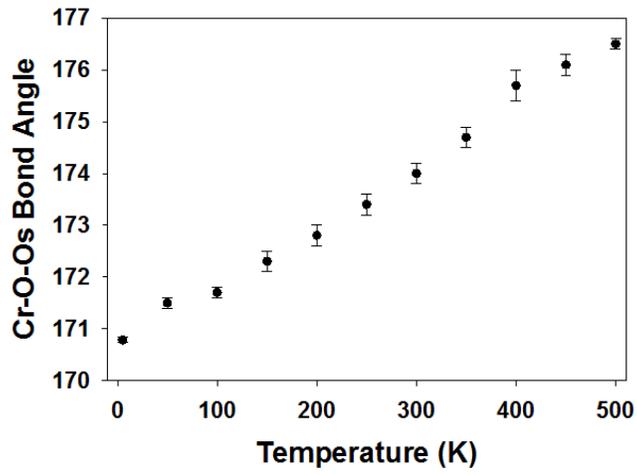

FIG. 3 (color online). The temperature dependence of the refined Cr−O−Os bond angle in $Sr_2CrOsO_6$.

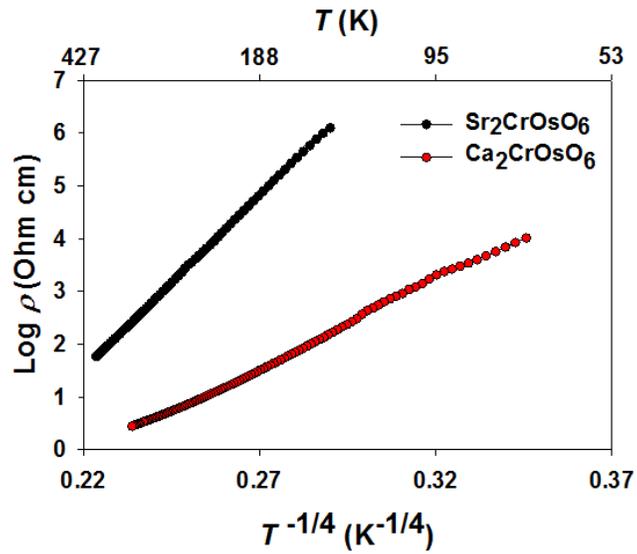

FIG. 4 (color online). The temperature dependence of the electrical resistivity of $Sr_2CrOsO_6$ (black) and $Ca_2CrOsO_6$ (red).

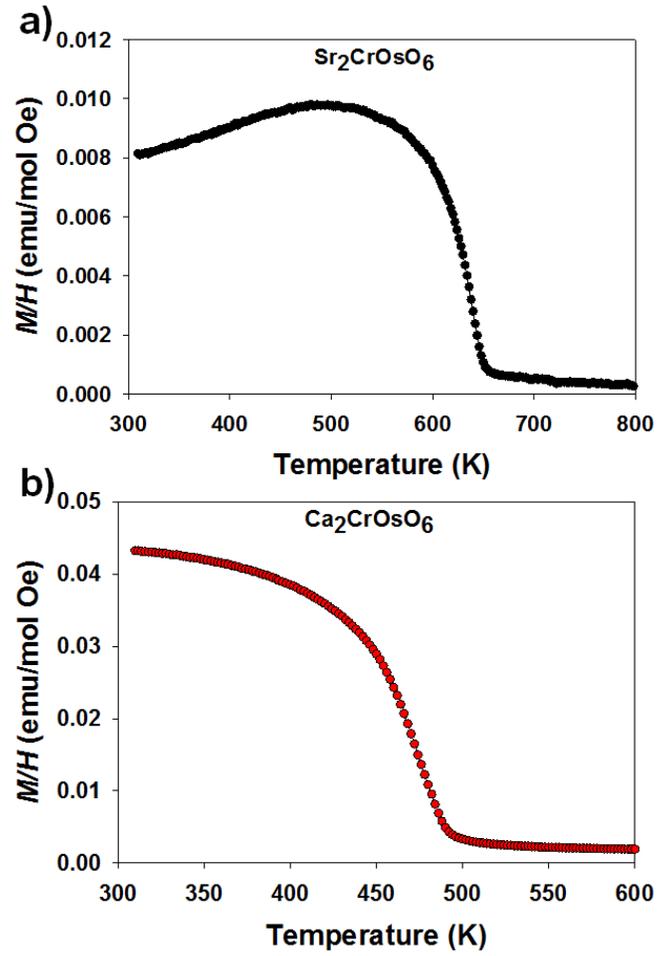

FIG. 5 (color online). The field cooled (15 kOe) temperature dependence of the magnetization of a) $Ca_2CrOsO_6$ and b) $Sr_2CrOsO_6$ measured under an applied field of 1 kOe.

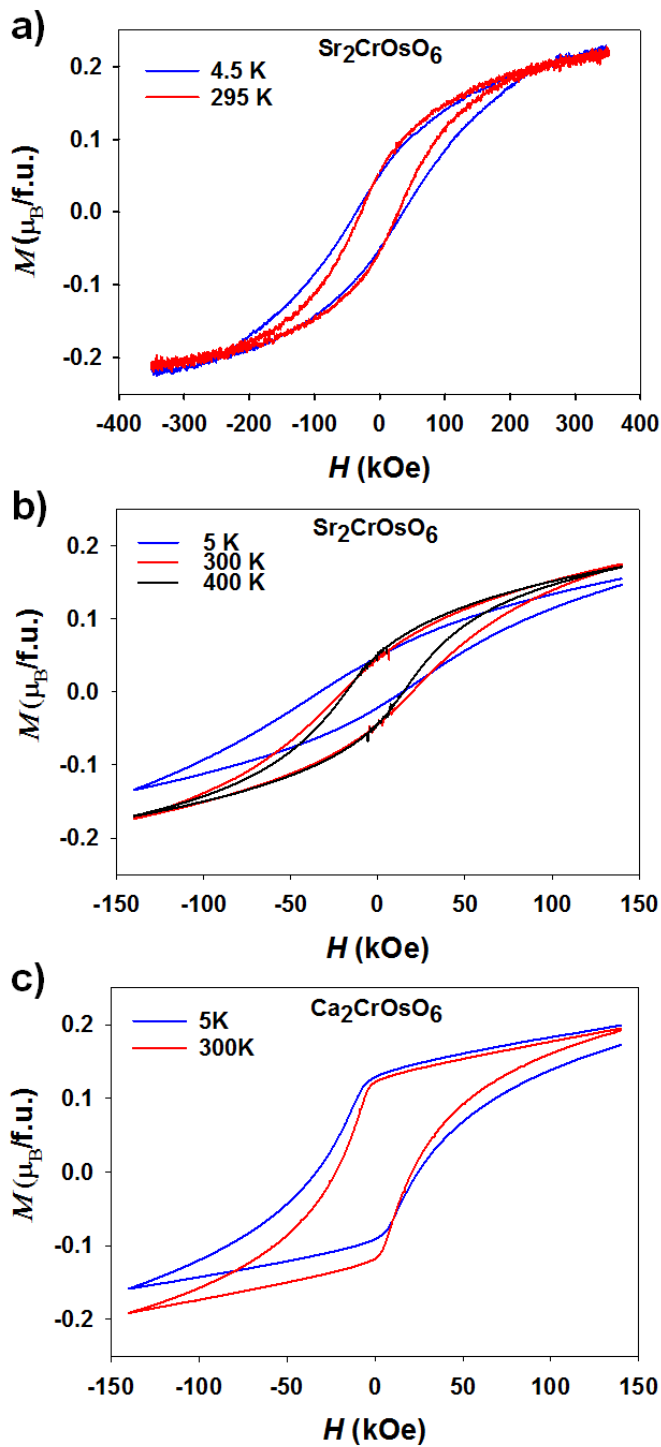

FIG. 6 (color online). The field dependence of the magnetization of a) $Sr_2CrOsO_6$ at 4.5 and 295K under a maximum applied field of 350 kOe as well as at b) 5, 300, and 400 K under maximum applied fields of 140 kOe. The field dependence of the magnetization of $Ca_2CrOsO_6$ at 5 and 300 K under a maximum applied field of 140 kOe is given in c).

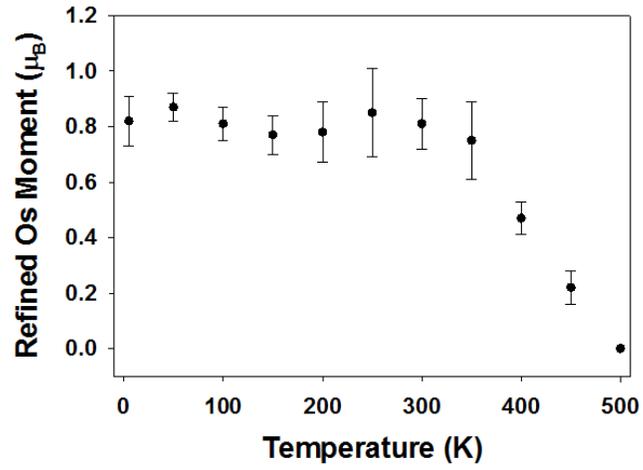

FIG. 7. The temperature dependence of the refined Os moment in $Sr_2CrOsO_6$ with a fixed Cr moment at 2.5 $\mu_B$ as described in the text. At 500 K, the Os moment is fixed to 0, and the Cr moment is refined to 2.24(6) $\mu_B$.

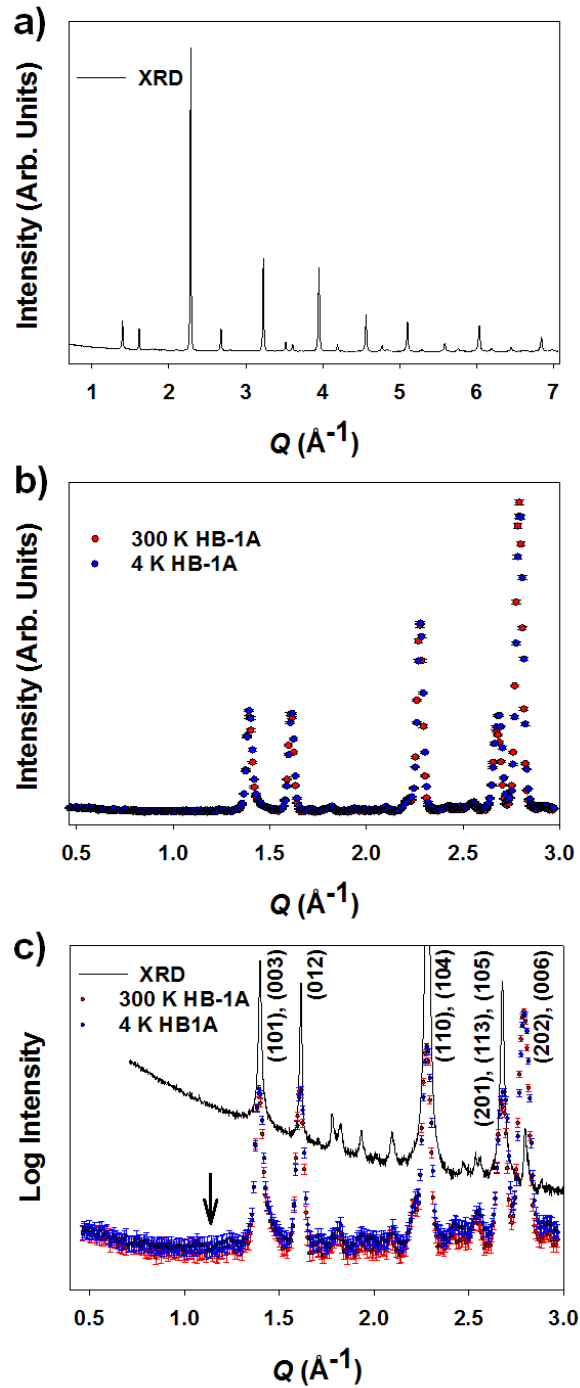

FIG. 8 (color online). a) Powder X-ray diffraction pattern of $Sr_2CrOsO_6$ sample used for HB-1A measurements, b) neutron scattering data from $Sr_2CrOsO_6$ powder collected at 4 and 300 K, and c) comparison of the data sets in log scale. The XRD pattern has been offset for clarity. All weak features seen in neutron scattering can be accounted for in the X-ray data indicating that they are not of magnetic origin and are likely associated with trace unknown impurities. The position of a predicted magnetic reflection as described in the text is indicated with an arrow.